# BlessMark: A Blind Diagnostically-Lossless Watermarking Framework for Medical Applications Based on Deep Neural Networks

Hamidreza Zarrabi[1], Ali Emami[1], Pejman Khadivi[2], Nader Karimi[1], and Shadrokh Samavi[1]

[1]Department of Electrical and Computer Engineering, Isfahan University of Technology, Isfahan, Iran
[2] Department of Computer Science, Seattle University, Seattle, WA 98122 U.S.A

*Abstract*— Nowadays, with the development of public network usage, medical information is transmitted throughout the hospitals. The watermarking system can help for the confidentiality of medical information distributed over the internet. In medical images, regions-of-interest (ROI) contain diagnostic information. The watermark should be embedded only into non-regions-of-interest (NROI) to keep diagnostic information without distortion. Recently, ROI based watermarking has attracted the attention of the medical research community. The ROI map can be used as an embedding key for improving confidentiality protection purposes. However, in most existing works, the ROI map that is used for the embedding process must be sent as side-information along with the watermarked image. This side information is a disadvantage and makes the extraction process non-blind. Also, most existing algorithms do not recover NROI of the original cover image after the extraction of the watermark. In this paper, we propose a framework for blind diagnostically-lossless watermarking, which iteratively embeds only into NROI. The significance of the proposed framework is in satisfying the confidentiality of the patient information through a blind watermarking system, while it preserves diagnostic/medical information of the image throughout the watermarking process. A deep neural network is used to recognize the ROI map in the embedding, extraction, and recovery processes. In the extraction process, the same ROI map of the embedding process is recognized without requiring any additional information. Hence, the watermark is blindly extracted from the NROI. Furthermore, a three-layer fully connected neural network is used for the detection of distorted NROI blocks in the recovery process, to recover the distorted NROI blocks to their original form. The proposed framework is compared with one lossless watermarking algorithm. Experimental results demonstrate the superiority of the proposed framework in terms of side information.

## I. INTRODUCTION

Transmission of Electronic Patient Record (EPR) between medical research centers, schools, and hospitals is necessary for educational purposes and e-healthcare applications such as teleconsulting, telediagnosis, remote surgery, etc. EPR encompasses confidential medical information about the patient, medical situation, diagnosis, and treatment [1]. Since EPR is transmitted through a public network such as the internet, the confidentiality protection of EPR is a real concern in the Health Information System (HIS). Watermarking is one solution to the confidentiality protection problem in HIS. Besides watermarking, various other methods have been proposed for improving the confidentiality protection of EPR. One approach is to embed encrypted EPR into a cover image with encryption algorithms such as chaotic encryption [2] and compressive sensing [3]. Another scheme is to embed the EPR into selected areas of the cover image rather than the whole cover image and use the binary location map as a key for embedding [4-5]. In the extraction process, the binary key used for the embedding process is required to extract the embedded EPR.

In the watermarking systems, capacity and imperceptibility are two vital concerns [6]. In other words, one concern in the watermarking schemes is to maximize embedding capacity, while the watermarked image suffers subtle distortions after embedding, i.e., the watermarked image would be perceptually indistinguishable from the original cover image. Another concern in watermarking systems is the concept of blindness [6]. Blind systems have been thoroughly investigated [7-10], and they generally lead to higher complexity. However, blind systems are practically superior to non-blind systems, which require various side information for the extraction or recovery on the receiver side [11-13].

Due to the rapid growth of machine learning and artificial intelligence in the last decades, several research works have used machine-learning tools for watermarking [14-19]. Some researches propose prediction based watermarking methods for natural images [14-15]. In natural images, pixels have a high correlation with their neighbors and can be predicted based on neighbor values. The prediction process may utilize various machine-learning tools such as the Extreme Learning Machine (ELM) [14] and Lagrangian Support Vector Regression (LSVR) [15]. Heidari et al. [16] embed the watermark data redundantly in multiple spectral zones. For extraction, they use SVM to detect a zone with minimum distortion, from which watermark is extracted. Abdelhakim et al. [17], use K-Nearest Neighborhood (K-NN) regression to find the optimum value of the embedding parameter. Among machine learning tools, deep Convolutional Neural Networks (CNN) have become popular in many computer vision applications such as object detection [20], pattern recognition [21], image classification [22], and watermarking [18-19]. For instance, Kandi et al. [19] use auto-encoders for feature extraction. In their method, watermark data is embedded in the feature space, generated by the auto-encoder, rather than embedding in a transform domain such as DCT.

Medical images are divided into two regions, ROI and NROI. Since the ROI region includes critical information for diagnostic purposes, small distortions of ROI may cause problems in medical diagnosis. On the other hand, NROI contains nonessential medical information. Hence, small deformations of NROI are tolerable and do not cause serious issues. In watermarking systems introduced by [13], [16], and [17], the whole cover image is considered for embedding, and this distortion caused during embedding is irreversible. Two approaches for medical watermarking are utilized to assure that diagnostic information is not distorted during watermarking or undesired distortion can be restored. The first approach is ROI-based watermarking, in which the cover image is divided into two regions, ROI and NROI. Then the watermark is embedded only in NROI. Thus, ROI remains intact, and only NROI is distorted during embedding. The second approach is the lossless watermarking. In this approach, the whole cover image is considered for embedding with lossless methods, so that in the recovery process, the original cover image is recovered without any loss of information.

One limitation of ROI based watermarking approaches is that they generally cannot detect the ROI map on the receiver side. Consequently, the ROI map needs to be sent as side information to the receiver side for the extraction of the watermark [23-24]. Hence, regardless of their segmentation method, they are non-blind algorithms, which require side information for the extraction of watermark data or the recovery of the original cover image. On the other hand, some other researches [25-27] do not send ROI map as side information. Therefore, they are blind algorithms, and the watermark data can be extracted solely from the watermarked image. Yang et al. [26] and Chaitanya et al. [27] embed ROI map inside the watermarked image to make it blind. However, the watermarking capacity is reduced as the embedded ROI map occupies part of data capacity. Perhaps the most relevant research to our work is [28], which can recognize the ROI map from the watermarked image on the receiver side and do not need to send ROI map as additional information. They segment the image by Otsu and then utilize a histogram-shifting method for the embedding of the watermark. This embedding method processes pixel values that are close to the peak bin. Hence, the ROI map is not changed during the embedding process and can be recognized before the extraction process. Despite the elegant idea suggested in this work, it is only applicable to the specific watermarking method utilized in this work. Hence, it cannot be generalized for other watermarking approaches. Furthermore, the proposed method suffers from insufficient evidence and experimental evaluation, as it is only testes on three medical image samples.

In lossless watermarking approaches, the whole cover image is distorted during embedding, since all regions of the image are considered for embedding. However, due to utilizing lossless methods, the original cover image is entirely recovered after watermark extraction [4], [29-35]. Despite complete recovery and extraction in the receiver side, some papers are non-blind, implying that some side information from the embedding module is required by the extraction or recovery modules on the receiver side [4], [31-33], [35].

In this paper, we propose a recursive method for blind diagnostically-lossless watermarking in medical applications, which is named BlessMark. We use a deep network for segmentation and generation of ROI map. Watermark is only embedded in NROI blocks within a novel iterative scheme. Therefore, sensitive medical information remains intact, which leads to a diagnostically-lossless watermarking system, where only NROI is distorted during the embedding process. The proposed framework is blind, as neither of the ROI map or the original cover image and information about watermark is transmitted to the receiver side. The utilized segmentation procedure helps further improve the confidentiality of the embedded information, due to the proprietary network weights, which are not publicly known by others. In other words, the ROI map may be considered as a key for confidentiality, since watermark data is extracted solely from NROI blocks.

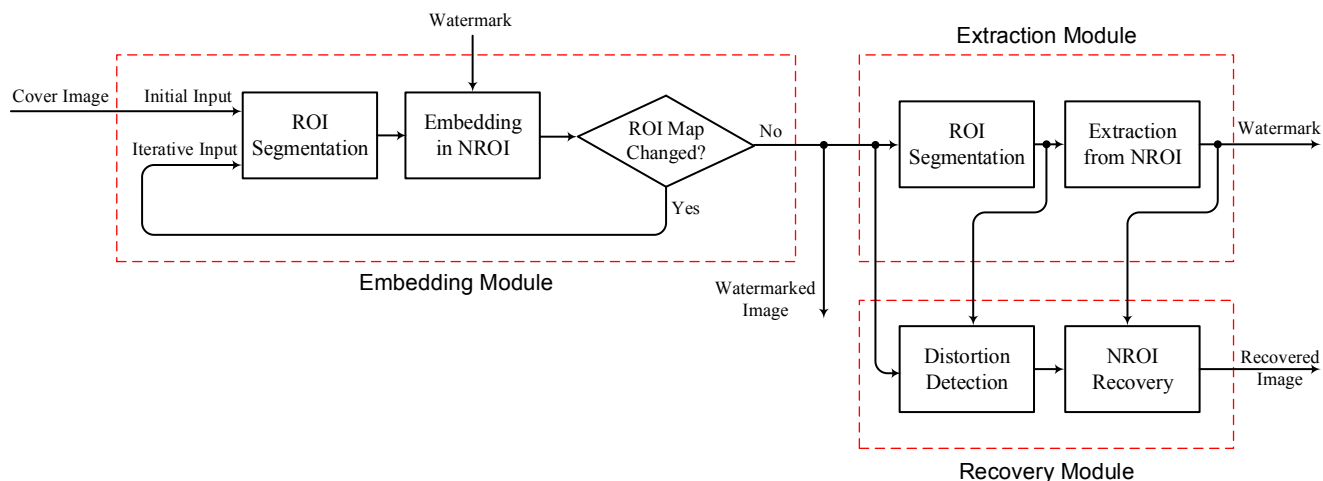

**Fig. 1** Block diagram of the proposed blind ROI-based watermarking framework.

Furthermore, we train a classifier to detect distorted NROI regions. Hence, distorted regions of NROI are mostly recovered to the original cover image.

To test our proposed framework, we use a simple embedding method in DCT domain. The choice of the DCT domain for embedding is just a convenient option for the proof of concept. Hence, different embedding methods in other transform domains, such as DWT and Hadamard, may be applied. Also, to evaluate the framework, we use a CNN for ROI segmentation and a three-layer fully connected neural network for the detection of distorted NROI blocks.

The main contributions of this work are as follows:

1) Introducing a framework that can be used as a platform for applying a desirable watermarking method to any medical image by utilizing various network structures in different transform domains. 2) In the extraction process, the same ROI map, which is used for the embedding process, is generated without any additional information. Therefore, the watermark is blindly extracted from NROI blocks. 3) NROI blocks, which are modified due to the embedding process, are recovered to their original state wherever possible.

The remainder of the paper is organized as follows: Details of the proposed framework is explained in Section II. The experimental results are explained in Section III. Finally, we conclude our proposed framework in part IV.

## II. GENERAL WATERMARKING FRAMEWORK

In this section, we go through the technical details of the proposed watermarking framework, BlessMark. As shown in the block diagram of Fig. 1, the framework is composed of three separate modules for the embedding of the watermark, extraction of the watermark, and the recovery of the original cover image. There is a standard core block, ROI segmentation network, which is responsible for segmentation the ROI regions and generating the ROI map for the embedding, extraction, and recovery modules. The problem is that embedding of the watermark data into the image may cause distortions, which could affect the segmentation results. Since we embed watermark data into NROI, the ROI map is critical for the extraction, and it is crucial in attaining the same segmentation on a watermarked image for accurate extraction of watermark data. To this end, we propose a novel iterative scheme for embedding based on the ROI segmentation function. This is done such that the ROI segmentation network can accurately detect the same ROI map in both transmitter and receiver sides. ROI map is automatically recognized on the receiver side without requiring any additional information. Therefore, the ROI map is key for confidentiality, as third parties cannot recognize it without having access to our proprietary segmentation tool.

On the other hand, due to embedding watermark data in NROI, crucial medical information in ROI remains intact. Another core block of the framework is a distortion detection network used in the recovery module. This network is responsible for the detection of distorted blocks due to the embedding process for the recovery operation in the next step.

In the following sections, we elaborate on the details of each block separately. We first introduce the proposed ROI segmentation network. In sections II-B and II-C, embedding and extraction modules are described. Details of the distortion detection network and the recovery module are presented in Section II-D.

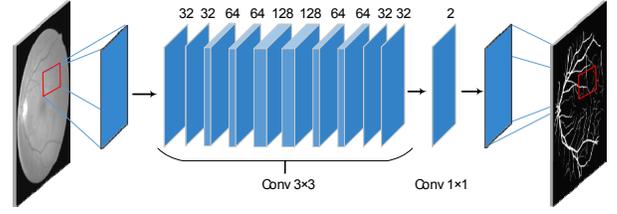

**Fig. 2** ROI segmentation network structure. Each blue box shows a multichannel feature map. Numbers above the boxes represent the number of channels.

### A. ROI Segmentation network

The ROI Segmentation unit is responsible for the detection of ROI pixels from NROI pixels, and the generation of ROI block map based on the ROI detected pixels. A block is considered as NROI if all of its pixels are detected as NROI. Otherwise, the block is considered as ROI. In this work, we use a CNN structure inspired by U-Net [36], without max-pooling, up-sampling, and concatenation layers, for the segmentation of the ROI pixels. Fig. 2 demonstrates the structure of the utilized network. As shown in Fig. 2, ROI segmentation network is composed of 11 convolutional layers. Input to the network is a small $m \times m$ image block, while block size remains constant across all the convolution layers. Hidden layers consist of $3 \times 3$ convolutions followed by ReLU (Rectified Linear Unit) activation function. At the final layer, a $1 \times 1$ convolution is applied, followed by the same ReLU activation function. The number of channels in each layer is shown on the top of each layer in Fig. 2. The network output represents a segmentation probability map for pixels of the input block. This probability map is converted to a binary map by thresholding on 0.5.

### B. Embedding Module

We propose a novel iterative approach for the embedding process. The block diagram of the embedding module is shown in Fig. 3. The ROI block map of the cover image is generated based on the strategy discussed in Section II-A. Watermark is embedded into NROI blocks of the cover image. Then NROI and ROI blocks are merged to construct the tentative watermarked image. The ROI block map of the tentative watermarked image is then generated based on the strategy discussed in Section II-A. One image block might be detected as NROI before embedding, and the same block may be detected as ROI after the embedding. If the watermarking process causes a change in the ROI block map, then the modified cover image is constructed, and the watermark is embedded into the NROI blocks of the modified cover image. The embedding process may cause some of the NROI blocks of the original cover image to be identified as ROI. The modified version of such NROI blocks are placed in the cover image and from there on they will be considered as ROI. We repeat the embedding process into the modified cover image until the ROI block map remains unchanged, and the final watermarked image is produced. Thus, we do not need to send the ROI block map to the receiver side with the watermarked image, and the proposed framework is blind.

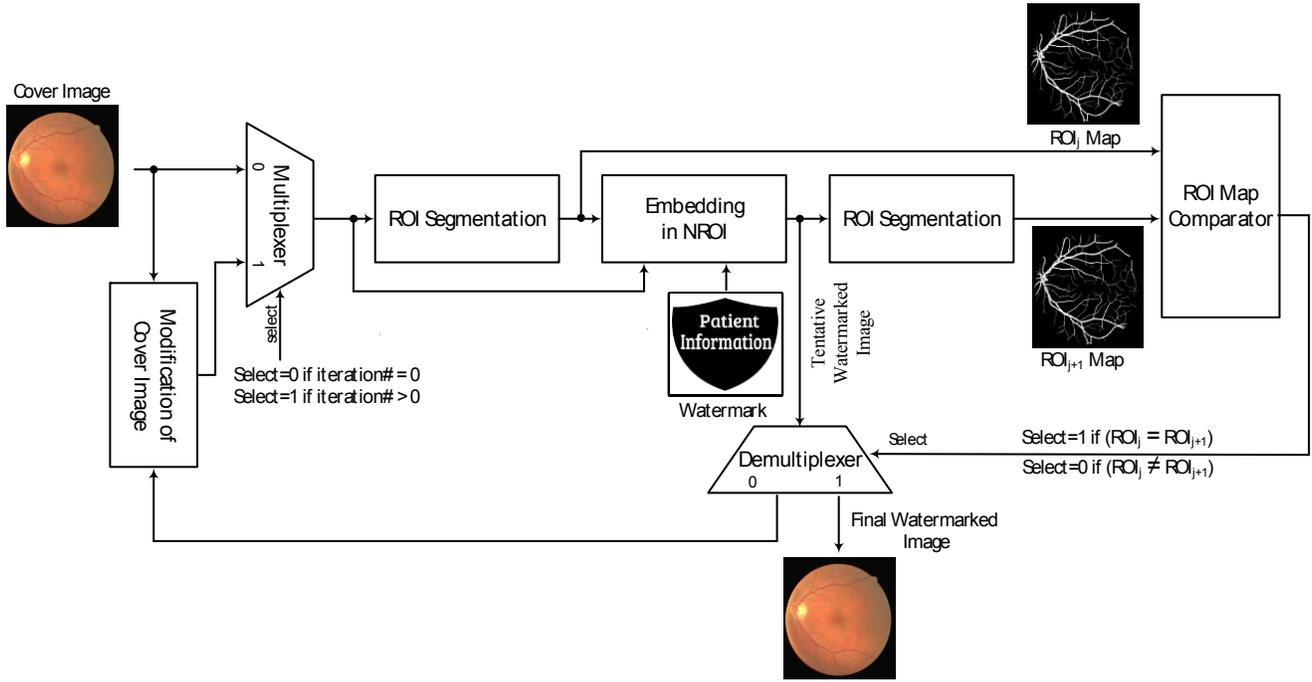

**Fig. 3** Block diagram of the BlessMark embedding module.

**Algorithm 1:** Embedding Pseudo Code
**Inputs:** Cover image and digital watermark ($w$)
**Output:** Watermarked image

**BEGIN**
01  $j = 1$
02  Segment cover image by CNN and generate ROI block Map, i.e. $ROI_j$
03  Apply DCT on non-overlapping NROI blocks
04  **IF** $w(k) == 0$
05    **IF** $c(i, i + 1) <= c(i + 1, i)$
06      $Swap(c(i, i + 1), c(i + 1, i)), c(i, i + 1) += th$
07    **ENDIF**
08  **ELSE**
09    **IF** $c(i, i + 1) >= c(i + 1, i)$
10      $Swap(c(i, i + 1), c(i + 1, i)), c(i + 1, i) += th$
11    **ENDIF**
12  **ENDIF**
13  Apply inverse DCT
14  Convert the output to integer
15  Merge NROI blocks with ROI blocks to construct a tentative watermarked image
16  Segment tentative watermarked image by CNN and generate ROI block map, i.e. $ROI_{j+1}$
17  **IF** $ROI_j \neq ROI_{j+1}$
18    Replace the switched NROI blocks of the original cover image by their embedded version to construct modified cover image
19    $j = j + 1$
20    Repeat steps 02 to 23 which modified cover image is considered as cover image in next iteration
21  **ELSE**
22    Consider tentative watermarked image as final watermarked image
23  **ENDIF**
**END**

We use a simple embedding method in DCT domain for proof of concept. However, the proposed framework is not limited to this specific domain, and watermarking can be performed in other known transform domains. A pseudo-code of the embedding module is presented in Algorithm 1. In the first step, $m \times m$ NROI blocks are detected. Then one bit of watermark is embedded in every NROI block. The embedding process is continued until the whole watermark is embedded. When two DCT coefficients are swapped, we add a constant threshold to secure a minimum distance between the two coefficients. Changing DCT coefficients may lead to under/overflow in the spatial domain, i.e., when they are transformed back to the spatial domain, the pixel values may exceed the valid range [0, 255]. In this situation, the under/overflowed pixels are assigned 0 and 255, respectively.

*C. Extraction Module*

In the extraction module, we blindly extract the watermark data from the watermarked image. The block diagram of the extraction module is shown in Fig. 4. The ROI map of the watermarked image is generated based on the strategy discussed in Section II-A. We accurately attain the same ROI map from the final embedding loop. The watermark is extracted from the NROI blocks of the watermarked image. The pseudo-code of the extraction module is presented in Algorithm 2. In the first step, the $m \times m$ NROI blocks are detected. Then one bit of watermark is extracted from each NROI block. The extraction process is continued until the whole watermark is extracted.

*D. Recovery Module*

The proposed recovery module performs recovery of distorted NROI blocks of the cover image to the extent feasible. We introduce a distortion detection classifier that is responsible

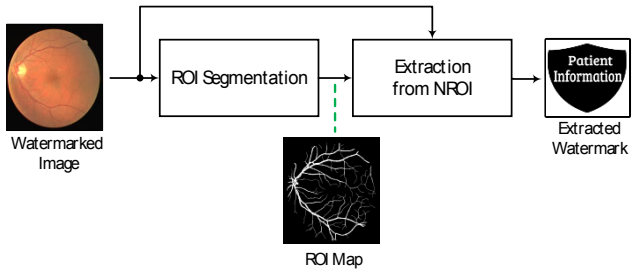

Fig. 4 Block diagram of the extraction module.

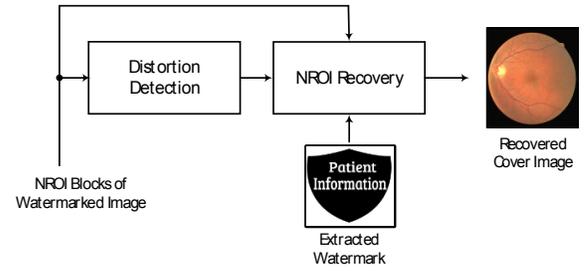

Fig. 5 Block diagram of the BlessMark recovery module.

---

**Algorithm 2:** Extraction Pseudo Code
**Input:** Watermarked image
**Output:** Extracted watermark ($w$)

**BEGIN**
01  Segment watermarked image by CNN and generate ROI block map, i.e. $ROI_j$
02  Apply DCT on non-overlapping NROI blocks
03  **IF** $c(i, i+1) >= c(i+1, i)$
04      $w(k) = 0$
05  **ELSE**
06      $w(k) = 1$
07  **ENDIF**
**END**

---

**Algorithm 3:** Recovery Pseudo Code
**Inputs:** Extracted watermark ($w$) and NROI blocks
**Output:** Mostly recovered NROI blocks of cover image

**BEGIN**
01  **IF** distortion detection network predict that NROI block has distorted
02      Apply DCT on block
03      **IF** $w(k) == 0$
04          $c(i, i+1) -= th$
05          **IF** $c(i, i+1) > c(i+1, i)$
06              $Swap(c(i, i+1), c(i+1, i))$
07          **ENDIF**
08      **ELSE**
09          $c(i+1, i) -= th$
10          **IF** $c(i, i+1) < c(i+1, i)$
11              $Swap(c(i, i+1), c(i+1, i))$
12          **ENDIF**
13      **ENDIF**
14      Apply inverse DCT
15      Convert the output to integer
16  **ENDIF**
**END**

---

for the detection of distorted NROI blocks that are altered during the embedding process from original NROI blocks. Then distorted blocks are recovered to the closest possible estimate of the original block. The block diagram of the recovery module is shown in Fig. 5. This module gets the NROI blocks of the watermarked image and the extracted watermark as its inputs and recovers the NROI blocks of the original cover image to the extent feasible. Recovered NROI blocks are merged with ROI blocks to construct the mostly recovered cover image. The pseudocode for the recovery module is presented in Algorithm 3. In the first step, $m \times m$ NROI blocks, which have been distorted through the embedding process, are detected. The distorted blocks are recovered to their original state by reversing the embedding operation. Underflow/Overflow pixels are assigned 0 and 255, respectively.

## III. EXPERIMENTAL RESULTS

The proposed framework is implemented in Tensor-flow [37] and executed on NVidia GeForce® GTX 1080 Ti. The framework is evaluated on two datasets of Retina [38] and X-ray Angiography [39], independently. The datasets are composed of 40 color images of size 584×565 and 44 grayscale images of size 512×512, respectively. We divided datasets into equal sets for training and test purposes. The CNN network and three-layer fully connected neural network performing ROI segmentation and distortion detection are separately trained on each dataset. We generate binary random sequences to be used as watermark data. For color images, watermark data is embedded in all three color channels. Thus, the distortion detection classifier is separately applied on single channels to detect the distorted blocks.

To evaluate the embedding method, we use the two standard measures, PSNR (Peak Signal to Noise Ratio) and SSIM (Structural Similarity Index). We utilize the Dice score to evaluate the segmentation network and the accuracy metric for evaluating the distortion detection classifier. Definitions of all the metrics are presented in Table 1. In the PSNR formula, $w$ and $h$ are the image dimensions. In this formula, $I$ and $I_W$ are the cover image and the watermarked image, respectively. In the SSIM formula, $I$ and $I_W$ are the cover and the watermarked images, respectively. In this formula, $\mu$ and $\sigma^2$ represent mean and variance values, respectively. Also, $\sigma_{I,I_w}$ is the covariance between $I$ and $I_W$ and $c_1$ and $c_2$ are constants. In Dice score and Accuracy measures, variables TP and FP are cases that are predicted as positive, while actual outputs are positive and negative, respectively. Also, TN and FN are cases that are predicted as negative, while the actual outputs are negative and positive, respectively.

All of the networks are evaluated in Section III-A. The performance of the whole framework is analyzed in section III-B. Finally, we compare BlessMark with state-of-the-art watermarking systems in section III-C.

**Table 1** Evaluation metrics employed for testing the framework.

| Evaluation | Metric | Formula |
|---|---|---|
| Watermark Imperceptibility | PSNR | $10 \, log_{10} \frac{255^2 \times w \times h}{\sum_{i,j=0}^{w,h}(I_W[i,j] - I[i,j])^2}$ |
| | SSIM | $\frac{(2\mu_I\mu_{I_w} + c_1)(2\sigma_{I,I_w} + c_2)}{(\mu_I^2 + \mu_{I_w}^2 + c_1)(\sigma_I^2 + \sigma_{I_w}^2 + c_2)}$ |
| Networks Evaluation | Dice Score | $\frac{2 \times TP}{2 \times TP + FP + FN}$ |
| | Accuracy | $\frac{TP}{TP + TN + FP + FN}$ |

**Table 2** Performance of segmentation network for two datasets with different block sizes.

| Block Size | | Test Dice Score |
|---|---|---|
| 6×6 | Retina | 0.73 |
| | Angio | 0.61 |
| 8×8 | Retina | 0.76 |
| | Angio | 0.67 |
| 10×10 | Retina | 0.76 |
| | Angio | 0.69 |

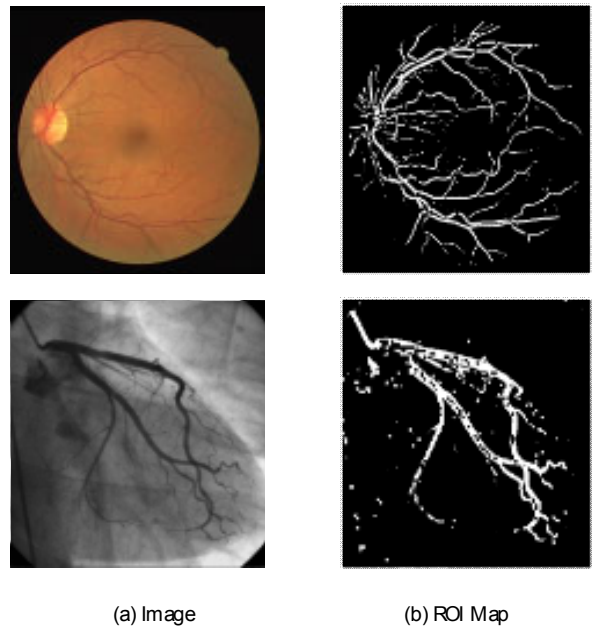

(a) Image     (b) ROI Map

**Fig. 6** ROI segmentation for Retina and Angiography image. (a) Image. (b) Produced ROI pixel map.

### A. Independent Evaluation of Networks

#### 1. Segmentation network

As discussed in previous sections, the segmentation network is responsible for attaining ROI pixel map. This network takes a block as the input and produces an ROI pixel map for that block. The network input is a gray-scale image block. Therefore, color images are converted to grayscale and one ROI map is generated for all channels. A softmax activation function is used for generating a probability map in the last layer. We use cross-entropy loss function with stochastic gradient descent (SGD) optimizer for training. Table 2 presents the Dice scores of the trained networks on two datasets with different block sizes. Two training sets constructed from 20 Retina images and 22 Angiography images are used separately for training each network within 150 epochs. The other half of the datasets are used as a test set. As shown in Table 2, the trained networks recognize 73% of ROI pixels for Retina and 61% of the ROI pixels for Angiography test sets, when the block size is 6×6. Dice scores increase by using larger block sizes, usually. Also, it is shown that a trained network for Retina attains a higher Dice score than a trained network for Angiography test sets in the same block size. As an example, the produced ROI pixel map for Retina and Angiography is shown in Fig. 6.

#### 2. Distortion Detection Network

As discussed in previous sections, the distortion detection network is responsible for detecting the distorted blocks, i.e., the blocks that have distorted during the embedding process. In this work, we train a three-layer, fully connected neural network for distortion detection. The structure of this network is shown in Fig. 7. The network input is a small $m \times m$ NROI block. In the first step, the block is flattened by the flattening layer. The two hidden layers are dense layers with size $m^2$ node, and each layer is followed by a ReLU activation function. At the final layer, a dense layer with one neuron is used, which is followed by a sigmoid activation function. We use Adam with default parameters as an optimizer and cross-entropy as a loss function. The network output represents a classification probability, in which this probability is converted to a binary by thresholding the output layer on 0.5.

Every block has an intrinsic embedded value. Hence, for arranging a training set for the network, we pick all the NROI blocks of all channels and invert the intrinsic embedded value inside them. The training set is composed of all the NROI blocks and their inverted versions. Thus, the three-layer fully connected neural network is trained to classify these two sets of blocks for all channels. It worth mentioning that we can embed different watermark data in various channels. Consequently, the classifier conducts independent analyses on separate channels.

The accuracy of the classifier for the two datasets is demonstrated in Table 3. Training set in the Retina and the Angiography datasets are used for the training of the two networks with 100 epochs. Test sets are constructed similar to the training sets by using a test set in the Retina and the Angiography datasets. Index $i$ in Table 3 represents the two DCT coefficients $c(i, i + 1)$ and $c(i + 1, i)$, which are swapped for embedding. As shown in Table 3, DCT coefficients swapped for embedding in various block sizes ('6×6', '8×8', '10×10') are different. The parameter $th = 0.01$ represents the threshold as used in our embedding algorithm (Algorithm 1). As shown in Table 3, the trained networks detect the distorted blocks with an accuracy of 94% for Retina and 97% for Angiography, when the block size is 6×6.

### B. Evaluation of the Whole Framework

In Table 4, we evaluate our framework for the 20 Retina and 22 Angiography test images in terms of imperceptibility and variations in the ROI map. Since the image is segmented by CNN, false classifications are probable, i.e., it is probable that one pixel is segmented as NROI by CNN while it is labeled as ROI in the ground truth or vice versa. Therefore, we use the ground truth ROI map for calculating PSNR of images.

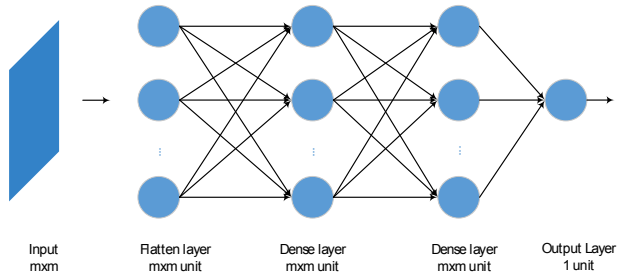

**Fig. 7** Distortion detection network structure.

**Table 3** Performance of classifier for two datasets with different block sizes. $th = 0.01$

| Block Size | | Index $i$ | Number of Blocks for Training | Number of Blocks for Testing | Test Accuracy |
|---|---|---|---|---|---|
| 6×6 | Retina | 5 | 820950 | 813042 | 0.94 |
| | Angio | 5 | 276894 | 277372 | 0.97 |
| 8×8 | Retina | 7 | 423726 | 418872 | 0.95 |
| | Angio | 7 | 155678 | 155736 | 0.97 |
| 10×10 | Retina | 9 | 248226 | 243858 | 0.95 |
| | Angio | 9 | 95524 | 95790 | 0.97 |

In Table 4, PSNR of the whole image, as well as PSNR of ROI and NROI regions, are presented. For Retina image with block size 6×6, PSNR of the watermarked image NROI is 54.96 (dB) on the maximum capacity of 0.021 (BPP). The image NROI imperceptibility is enhanced to 59.81 (dB) after recovery, i.e., imperceptibility of NROI may be improved by 4.85 (dB). The recovery process is not perfect due to the distortion detection classifier errors, irreversible distortions caused by under/overflow clipping of embedding operation, and switching of segmentation results after embedding. We also demonstrate the average percentage of NROI blocks, which are switched to ROI blocks during the embedding process. It is shown that distortions caused by iterative embedding process are minimal and decrease by using larger block sizes.

Boxplot of Fig. 8 shows the distribution of PSNR and SSIM for watermarked images across the Retina and Angiography test sets. Each box corresponds to block size. For block sizes 6×6, 8×8, 10×10, maximum experimental capacities are 0.021, 0.010, and 0.006 (BPP) for Retina and 0.022, 0.012 and 0.007 (BPP) for Angiography test sets.

In Table 5, the training durations for the segmentation and classifier networks are shown for the Retina training set with $6 \times 6$ block size. The segmentation network is trained by 2M blocks in 150 epochs, and the classifier network is trained by 820950 blocks in 100 epochs. The segmentation and the classification networks are trained in 23 hours and 4.5 hours, respectively.

In Table 6, embedding, extraction, and recovery times with GPU and without GPU are evaluated for the Retina test set. Average experimental results for 20 Retina test images with block size $6 \times 6$ and capacity 0.021 (BPP) is demonstrated in Table 6. Embedding time with GPU is 7 seconds. Extraction and recovery time with GPU is 3 seconds. These times are increased to 78 and 18 seconds without GPU.

In Fig. 9, the visual qualities of the embedding result with block size $6 \times 6$ for two test images are shown. The left column shows the original cover image, the middle column is the embedded watermark, and the right column demonstrates the watermarked image with its corresponding PSNR and SSIM. We have not seen any distinguishable difference between the watermarked image and the cover image.

### C. Comparison with State-of-the-art Methods

Properties of various watermarking algorithms [4], [23-24], [26-28], [29-35] are compared with our framework in Table 7. The eight lossless watermarking methods [4], [29-35] use the whole cover image for embedding, causing distortions across all image regions, including ROI. Therefore, they are not applicable to medical watermarking, since diagnostic information in ROI may be corrupted. Also methods of [4], [31-33], [35] are non-blind.

The two non-blind ROI based watermarking methods [23-24] send ROI map with the watermarked image to the receiver side. The two blind ROI based watermarking methods [26-27] embed ROI map inside the watermarked image. Hence, the extraction of ROI map from the watermarked image is a prior step for the extraction of watermark data. The ROI based watermarking method [28] utilize the same segmentation method for recognizing ROI map in embedding and extraction modules of their proposed system. In our framework, the watermark is embedded in NROI blocks to keep the sensitive ROI information intact during the embedding process. ROI map can be recognized from the watermarked image in the extraction module by the trained segmentation network.

**Table 4** Performance of the proposed framework on Retina and Angiography test sets.

| Block size | | Capacity (BPP) | PSNR of Watermarked Image (dB) | PSNR of Recovered Image (dB) | Improvement of PSNR (dB) | PSNR of Watermarked Image NROI (dB) | PSNR of Recovered Image NROI (dB) | Improvement of NROI PSNR (dB) | PSNR of Watermarked Image ROI (dB) | PSNR of Recovered Image ROI (dB) | Improvement of ROI PSNR (dB) | Average Percent of Switched NROI Block into ROI Block (%) |
|---|---|---|---|---|---|---|---|---|---|---|---|---|
| 6×6 | Retina | 0.021 | 55.24 | 60.09 | 4.85 | 54.96 | 59.81 | 4.85 | 60.78 | 65.29 | 4.51 | 0.164 |
| | Angio | 0.022 | 55.65 | 62.81 | 7.16 | 55.48 | 62.66 | 7.18 | 59.63 | 65.96 | 6.33 | 0.047 |
| 8×8 | Retina | 0.010 | 58.16 | 63.16 | 5.00 | 57.84 | 62.84 | 5.00 | 65.31 | 69.90 | 4.59 | 0.154 |
| | Angio | 0.012 | 58.20 | 65.32 | 7.12 | 57.99 | 65.11 | 7.12 | 63.73 | 70.88 | 7.15 | 0.040 |
| 10×10 | Retina | 0.006 | 60.77 | 64.50 | 3.73 | 60.44 | 64.17 | 3.73 | 68.50 | 72.64 | 4.14 | 0.137 |
| | Angio | 0.007 | 61.39 | 64.67 | 3.28 | 61.16 | 64.44 | 3.28 | 67.96 | 71.09 | 3.13 | 0.030 |

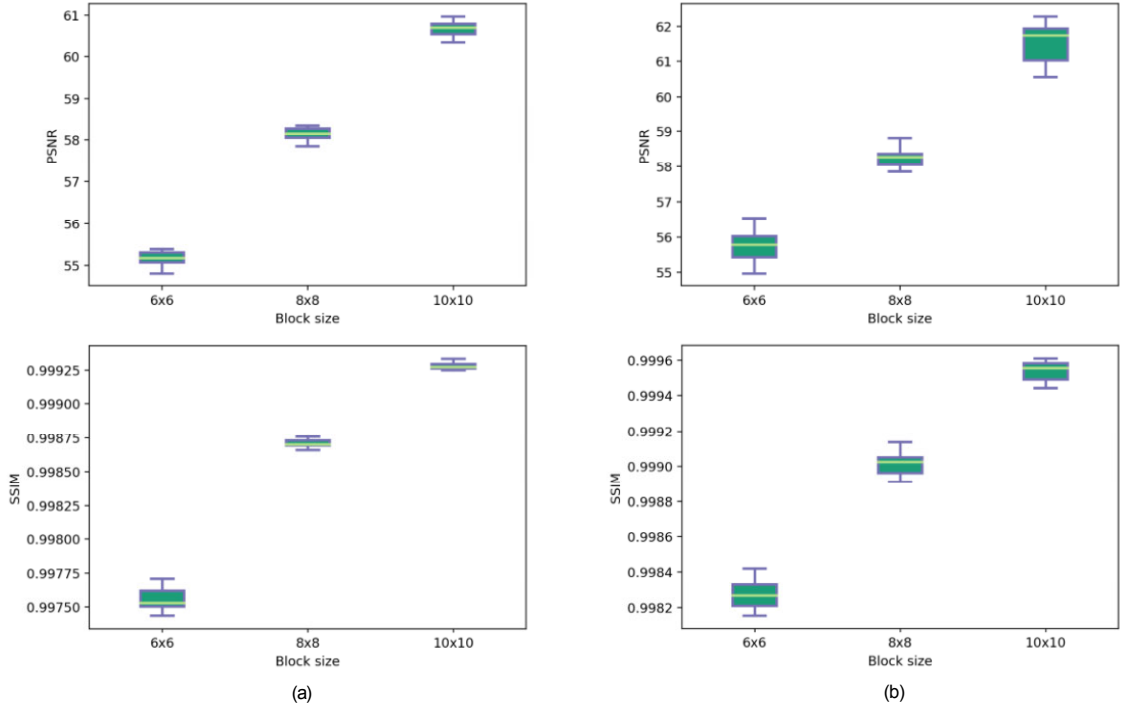

(a)                              (b)

**Fig. 8** Box-plot of watermarked image PSNR and SSIM. Each box represents the distribution of PSNR and SSIM versus block size. (a) Retina. (b) Angiography.

**Table 5** Training times of two networks on Retina training set.

| Segmentation network | Classification network |
|---|---|
| 23 (Hour) | 4.5 (Hour) |

**Table 6** Average times of the proposed framework on Retina test set.

| Time (Second) | CPU | GPU | RAM (GB) | Software |
|---|---|---|---|---|
| 7 + 3 | Core i7, 4.2 GHz | NVidia GeForce 1080 Ti | 64 | Spyder (Python) |
| 78 + 18 | Core i7, 1.6 GHz | No | 4 | Spyder (Python) |

In Table 8, imperceptibility and side-information of one lossless method [33] are compared with our system. Similar to Table 4, we use the ground truth ROI map for calculating PSNR of ROI and NROI. Since the false classification of CNN is probable, i.e., that one pixel may be segmented as NROI by CNN while it is labeled as ROI in the ground truth or vice versa.

The lossless method [33] embeds the watermark into 1-level of IWT of the whole image in two iterations so that iteration-2 compensate produced distortion in iteration-1.

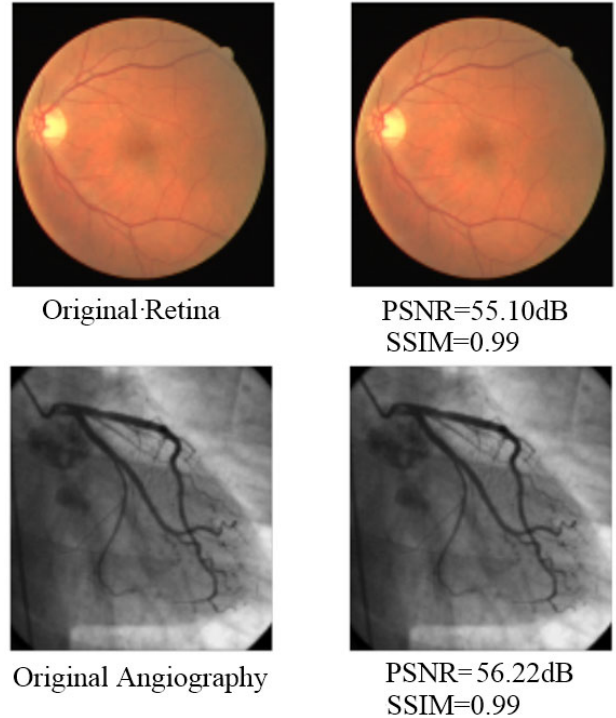

Original Retina        PSNR=55.10dB SSIM=0.99

Original Angiography        PSNR=56.22dB SSIM=0.99

**Fig. 9** (a) Original cover image. (b) Watermarked image.

**Table 7** Qualitative comparison of the proposed framework with other methods.

| Reference | Embedding Region | ROI Segmentation Method | ROI Map as Additional Information | Blind | Watermarking Approach |
|---|---|---|---|---|---|
| **Proposed** | **NROI** | **CNN** | **No** | **Yes** | **ROI-based** |
| [23] | NROI | Manual | Yes | No | ROI-based |
| [24] | NROI | Thresholding of saliency map | Yes | No | ROI-based |
| [26] | NROI and ROI | Adaptive Threshold Detector | Yes | Yes | ROI-based |
| [27] | NROI | Manual | Yes | Yes | ROI-based |
| [28] | NROI | Otsu | No | Yes | ROI-based |
| [4] | Whole | No segmentation | No | No | Lossless |
| [29] | Whole | No segmentation | No | Yes | Lossless |
| [30] | Whole | No segmentation | No | Yes | Lossless |
| [31] | Whole | No segmentation | No | No | Lossless |
| [32] | Whole | No segmentation | No | No | Lossless |
| [33] | Whole | No segmentation | No | No | Lossless |
| [34] | Whole | No segmentation | No | Yes | Lossless |
| [35] | Whole | No segmentation | No | No | Lossless |

**Table 8** Comparison of average PSNR and side information with other method for different block sizes.

| Reference | Block Size | Capacity (BPP) | PSNR of Watermarked Image (dB) | PSNR of Watermarked Image NROI (dB) | PSNR of Watermarked Image ROI (dB) | PSNR of Recovered Image (dB) | PSNR of Recovered Image NROI (dB) | PSNR of Recovered Image ROI (dB) | Average Side Information (Bit) |
|---|---|---|---|---|---|---|---|---|---|
| Proposed | 6×6 | 0.020 | 55.01 | 54.64 | 60.95 | 59.71 | 59.32 | 65.90 | **0** |
| | 8×8 | 0.010 | 58.43 | 58.02 | 65.84 | 62.17 | 61.77 | 69.19 | **0** |
| | 10×10 | 0.005 | 60.87 | 60.45 | 68.67 | 64.77 | 64.35 | 72.74 | **0** |
| [33] | Pixel based | 0.020 | 65.01 | 64.75 | 70.06 | Inf | Inf | Inf | 16516.5 |
| | | 0.010 | 67.99 | 67.67 | 74.63 | Inf | Inf | Inf | 8242.5 |
| | | 0.005 | 70.55 | 70.22 | 77.28 | Inf | Inf | Inf | 4627.5 |

In Table 8, the PSNR and side-information results are calculated over the 20 Retina test images. For a fair comparison, a single watermark is embedded in all methods. The infinity values of PSNR in the table demonstrate that the original cover image has been completely recovered. The proposed framework does not produce any side information, while the method of [33] produce 16516.5 bits of side information on the capacity of 0.020 (BPP). However, the lossless method [33] can completely recover the original cover image. Since NROI does not contain important information for medical diagnosis, accurate recovery of NROI is not critical in medical applications.

## IV. CONCLUSION

In this paper, we presented BlessMark, a framework for blind diagnostically-lossless watermarking. The proposed watermarking scheme is used as a means for the simultaneous improvement of confidentiality and preservation of diagnostic medical information. BlessMark consists of a deep neural network for image segmentation and one fully connected neural network for classification.

Its ROI segmentation network generates an ROI map for the embedding, extraction, and recovery modules. The ROI segmentation network is applied using an iterative scheme to accurately generate the same ROI map, both in the transmitter and receiver sides. Hence, in the proposed blind watermarking framework, an ROI map is automatically generated on the receiver side without requiring any additional information. The proposed ROI map is critical for improving confidentiality protection in our system, as third parties cannot generate it without having access to our proprietary segmentation tool. Since the watermark is embedded only in NROI blocks, ROI remains intact and leading to a diagnostically-lossless watermarking system. The distortion detection classifier used in the recovery module helps the detection of blocks that have distorted during the embedding process. Distorted blocks are mostly recovered to their original form.

The choice of a simple embedding method in the DCT domain is just a convenient option to prove the concept of the proposed framework. Hence, different embedding methods in other transform domains such as DWT and Hadamard may be applied. Furthermore, we used a CNN and a three-layer fully connected neural network for the ROI segmentation and detection of distorted NROI blocks. However, other structures can be investigated for these purposes. Our watermarking method is non-robust since the ROI map of the watermarked image can be changed as a result of attacks. Since the ROI map is vital for embedding and extraction, it is necessary to detect the same ROI map on both sides.

We evaluated our framework on two Retina and Angiography datasets. The proposed framework demonstrates producing any side information in comparison with one lossless watermarking method.

One future research is to explore the capabilities of the proposed framework by using it as a platform for testing other networks and embedding domains.